\documentclass{nature}
\usepackage{graphicx}
\usepackage{epsfig}
\usepackage{pdfpages}
\usepackage{color}
\usepackage{wrapfig} 
\usepackage{float} 
\usepackage[mediumspace,mediumqspace,Grey,squaren]{SIunits}

\title{Pressure and temperature dependence of growth and morphology of
  Escherichia coli: Experiments and Stochastic Model}

\author{Pradeep Kumar$^{1}$, and Albert Libchaber$^{1,2}$}

\begin{document}

\maketitle

\begin{affiliations}

\item Center for Studies in Physics and Biology, The Rockefeller
  University, 1230 York Avenue, New York, NY 10021 USA.

\item Institue for Advanced Study, Einstein Drive, Princeton, New
  Jersey 08540 USA.
 
\end{affiliations}
\begin{abstract}

We have investigated the growth of Escherichia coli \textit{E.coli}, a
mesophilic bacterium, as a function of pressure $P$ and temperature
$T$. \textit{E.coli} can grow and divide in a wide range of pressure
($1-400$atm) and temperature ($23-40^{\circ}$C). For $T>30^{\circ}$~C,
the division time of \textit{E.coli} increases exponentially with
pressure and exhibit a departure from exponential behavior at
pressures between $250-400$~atm for all the temperatures studied in
our experiments. For $T<30^{\circ}$~C, the division time shows an
anomalous dependence on pressure -- first decreases with increasing
pressure and then increases upon further increase of pressure. The
sharp change in division time is followed by a sharp change in
phenotypic transition of E. Coli at high pressures where bacterial
cells switch to an elongating cell type. We propose a model that this
phenotypic changes in bacteria at high pressures is an irreversible
stochastic process whereas the switching probability to elongating
cell type increases with increasing pressure. The model fits well the
experimental data. We discuss our experimental results in the light of
structural and thus functional changes in proteins and membranes.

\end{abstract}
\section*{Introduction}

A vast majority of bacteria and archaea can grow in diverse
environmental conditions. The range of those conditions include high
pressures~\cite{Yayanos1981,Kato1998}, high
temperature~\cite{Brock1969}, low temperature~\cite{Bakermans2003},
high salinity, low~\cite{Schleper1995} and high
pH~\cite{HorikoshiBook1982,Horikoshi2008} etc. Since these conditions
are not hospitable for other life forms hence these organisms are
named
extremophiles~\cite{Brock1969,Oshima1974,Yayanos1981,Hebraud1999,Sharma2002,Deguchi2011,Sakiyama1998}.
One of the first isolated extremophile {\it Thermus aquaticus}, a
thermophilic bacteria can survive at near-boiling
temperatures~\cite{Brock1969}. Adaptation of these organisms to such
harsh conditions raises many interesting questions--how do they adapt
to these conditions ? Does the adaptation occur at single component
level such as mutations in proteins leading to their barostability and
thermostability, or the adaptation to these conditions has a collective
nature --whereas more than one cellular components act in compliance
to preserve the functionality of each other.

\noindent Recent studies on various aspects - such as taxonomy,
ecology, enzymology of these microorganisms have provided insights on
the adaptation of these organism to their environmental
conditions~\cite{Hebraud1999,Sterner2001,Charlier2005}. For example,
the bacterial cytoplasmic membrane must maintain its
liquid-crystalline structure and semipermeability with changing
conditions~\cite{Kaneshiro1995}. It was shown that bacterial membrane
adapt to the temperature changes by changing their lipid
composition~\cite{Russell1983}. Adaptation of a protein to non-ambient
conditions requires that it maintains its catalytic activity as well
as its structure~\cite{Jaenicke1998,Razvi2006}. Most globular proteins
denature both at high as well as low temperatures. Moreover, even if a
protein does not denature at low temperature, small thermal
fluctuations will lead to decreased catalytic activity at low
temperatures. Indeed in few studies on proteins from psychrophilic
organisms, it was found that proteins are more
flexible~\cite{Peterson2007}. However, increase of flexibility also
leads to high propensity of unfolding of the protein. Hence a fine
balance between the structural flexibility and stability is
required~\cite{Feller2003}. Recent comparative study of an essential
recombination protein RecA from mesophilic and thermophilic bacteria
suggests that its function of binding to single stranded DNA is
adapted to the conditions in which organisms grow~\cite{Merrin2011}. A
study of SSB, a single-stranded DNA binding protein, from mesophilic
and piezophilic bacteria show similar adaptation~\cite{Chilukuri2002}.

\noindent While there is a large body of work on the stability and
kinetics of proteins and adaptation of different components of
prokaryotes obtained from extremophiles, the growth of bacteria is only
approached using conventional methods such as plate counting. Such
studies have provided killing curves of saturated bacterial solutions
upon increasing pressures and hence a pressure-temperature phase
diagram of the bacterial survival is obtained~\cite{Ludwig1992}. 

\noindent In order to understand the adaptation of bacterial cells to
extreme pressure and temperature, an understanding of growth
bottlenecks and physical changes of bacterial growing at ambient
conditions induced by different thermodynamic conditions is
important. In this paper, we study the pressure-temperature dependence
of growth and phenotypic changes of a mesophilic bacterium, {\it
  E. coli}, using an optical method which allows us to measure the
growth of bacterial in real time at different pressures and
temperatures. We have investigated the growth and morphological
changes in a wide range of pressure and temperature. In ``Methods'',
we describe our experimental setup and protocol to measure growth of
bacteria. In ``Results'', we summarize the results of the
pressure-temperature dependence of growth followed by a stochastic
model to account for the morphological changes induced by high
pressure. Finally, we summarize and discuss our results in ``Summary
and Discussions''.

\begin{methods}

\section*{Experimental Setup}

\noindent Measurement of growth at normal conditions is rather easy as there are
many commercial photometers available. High pressure and temperature
require that a photometer optics is built around a high pressure cell
in order to obtain the growth curve. Below we describe our
experimental setup to measure the growth.

\noindent Bacteria absorb and scatter light with intensity which
depends on the scattering angle and absorption
coefficient~\cite{BohrenBook}. The most common method of measuring
bacterial concentration in a solution is turbidity method where the
extinction of the light is measured at a fixed angle, usually in
forward direction.  The method relies on many assumptions including
(i) each bacterial cell is an independent scatterer (ii) the shape of
bacterial cell is uniform (iii) multiple scattering is negligible.
The extinction cross section $C_{ext}$ is a sum of cross section due
to scattering $C_{sca}$ and due to absorption $C_{abs}$, and can be
written as
\begin{equation}
C_{ext} = C_{sca} + C_{abs}
\end{equation}
Then the coefficient of extinction $\alpha$ is
\begin{equation}
\alpha = \rho C_{ext}
\end{equation}
where $\rho$ is the number density of bacterial cells. The intensity
$I_t$ detected by a light detector after the incident light traverses
a distance $x$ in the scattering medium is given by
\begin{equation}
I_t = I_0 e^{-\alpha x}
\end{equation}
where $I_0$ is the intensity of the light incident on the
medium. Hence the difference of the logarithm of the intensities of
incident and the transmitted light is proportional to the
concentration of scatterers in the medium. The optical density (OD) of
the medium is thus defined as
\begin{equation}
OD = log(I_0/I_t)
\end{equation}

\noindent A schematic of our experimental setup to measure pressure
temperature dependence of growth is shown in Fig.~\ref{fig:fig1}. 

\begin{figure}[!htb]
\begin{center}
\includegraphics[width=8cm]{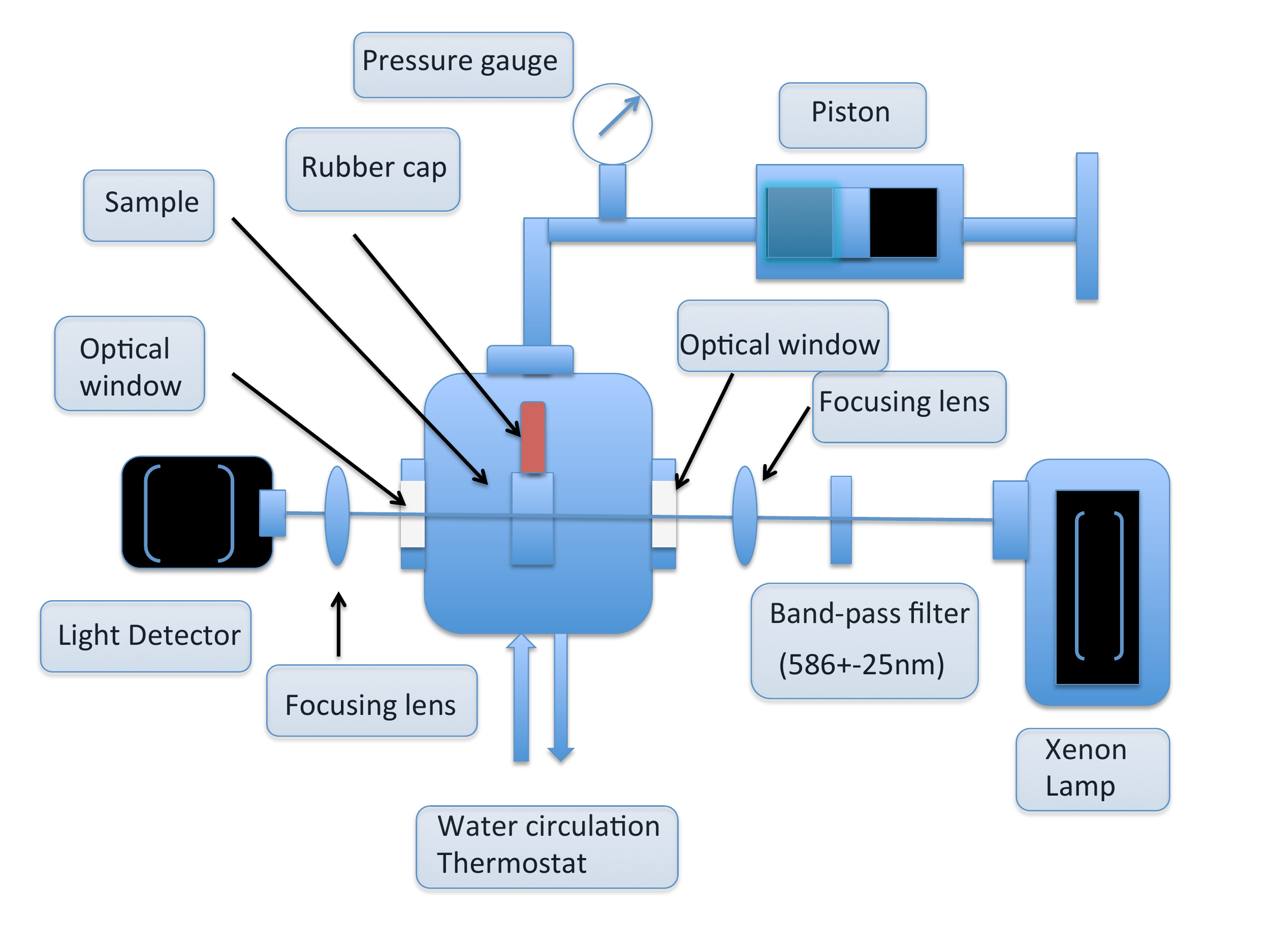}
\end{center}
\caption{Schematic of the experimental setup to measure bacterial
  growth.}
\label{fig:fig1}
\end{figure}
A sample of bacteria with LB medium is contained in a rectangular
cuvette (Spectrocell; volume: 400$\mu L$) made up of fused silica and
having a square cross-section (6mmx6mm). The cuvette with a flexible
movable teflon cap is loaded into the high pressure cell (ISS,
Illinois, USA). A piston is used to pressurize the water inside the
pressure cell, and the pressure is measured by a pressure gauge. The
growth of the bacteria is measured in real time by shining a white
light (xenon lamp) which passes through an excitation bandpass filter
(Semrock, FF01-586/15-25) and is focused onto the cuvette holding the
sample. We chose light with $586$~nm wavelength to keep our
measurements consistent with measurements done with most of the
commercially available photometers. The transmitted light is focused
on a light detector on the other side of the cuvette which measures
the intensity of the transmitted light. The light detector is built
around photo sensor chip TSL230R (TAOS). TSL230R photo sensor consists
of a silicon photodiode with a current to frequency converter built
into it. The nonlinearity error is typically 0.2\% at 100kHZ. The
frequency of the TTL signal from the detector proportional to the
incident light intensity is measured using MIO16 frequency counter
chip interfaced to Labview software (National Instruments). We
maintain the intensity of the light source such that all our
experiments fall into the linear regime of the sensor. The distance
between the light sensor and the cuvette was $10$~cm. Distance between
the detector and the sample dictatates the angular integration of the
scattered light incident on the sensor. The temperature of the high
pressure cell is regulated using a circular water bath thermostat. The
time for growth measurement ranged between $500$ and $1000$~mins,
depending on the pressure-temperature dependent growth rate of
bacteria. The entirely closed structure of our experimental setup
imposes a major limitation on the regulation of oxygen in our
experiments. The growth measurements were done in oxygen limited
conditions. The partial pressure of oxygen in LB medium was $20$kPa.

\section*{Cell culture and growth medium}

\subsection{\bf\textit{Bacteria and Media.}} 

For all the experiments reported here, DH$5\alpha$ strain of
\textit{E. coli} was used. While other wild type strains of E. Coli
such as MG1655 (K-12) or MC1000 are common for studying the physiology
of bacterial cells due to least amount of genetic mutations,
DH5$\alpha$ offers certain advantages for our studies. Earlier studies
have shown that a major effect of the pressure on the morphology is
elongation of cells. Due to cell elongation, SOS system is implicated
in the change of morphology at high pressures. The recA1 mutation in
DH5$\alpha$ causes the elimination of the homologous recombination, an
initiatiator process for SOS pathway upon UV
irradiation~\cite{Michel2005}. Lack of RecA mediated recombination in
DH$\alpha$ removes the effect of pressure on the SOS pathway. Hence
our experimental results would be able to distinguish the high
pressure effects where recombination system is not involved (discussed
in the Summary and Dicussion section). The drawback of using
DH$5\alpha$ is that since it lacks the homologous recombination
system, the cell are sickly and the growth is slower compared to other
wild type strains. Due to its slow growth, cells were grown in
standard Luria Broth (LB) medium~\cite{LB}, which is a rich medium for
bacterial growth. The pH of the growth medium was kept to $7$ by
adding NaOH to the solution. For the consistency of the experiments,
cells were first grown on a LB plate for about 10 hours and then
subsequently used for experiments as described below.

\noindent {\bf \textit{Growth conditions and measurements.}} Bacterial
cells picked from LB plate were first grown in LB medium at
atmospheric pressure and $T=37^{\circ}$C in an incubator until the
optical density (OD) of the solution is about $1.0$. A small amount of
freshly grown bacterial cells was then added to a cuvette containing
$800 \mu$~L of medium to bring the initial OD to $0.005$ and was used
as the starting point for all the pressure temperature
measurements. The final bacterial solution with LB medium was then
transfered to a high pressure cuvette at room temperature and pressure
and was closed with a teflon cap. The cuvette with the bacterial
solution was then put into the high pressure chamber (see the
experimental setup) equilibrated at the temperature of interest and
the piston of the high pressure setup was slowly increased until the
pressure gauge reading reaches the desired value of the pressure. The
growth of the bacterial cells then was assessed by measuring
extinction of light as described in our experimental setup. Growth
measurements were done in a sealed high pressure cell. Images were
taken using Sensicam cooled CCD camera connected to Zeiss Axiovert 35
microscope with a 40X Olympus objective. Image analysis of bacterial
cells was done using ImageJ software~\cite{ImageJ}.

\end{methods}

\section*{Results} 

\section*{Exponential dependence of division time with pressure}

\begin{figure}[!htb]
\begin{center}
\includegraphics[width=8cm]{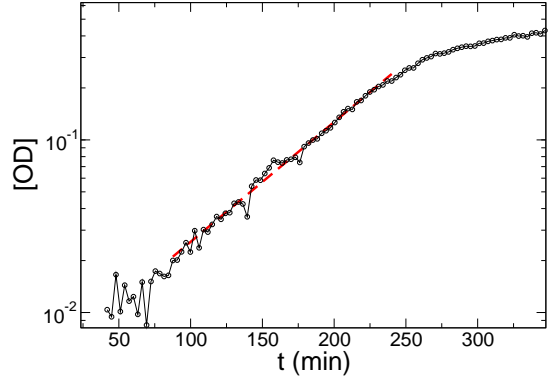}
\end{center}
\caption{Growth curve at $P=1$~atm and $T=37^{\circ}$~C. The growth
  shows a typical lag, exponential, and saturation regimes}
\label{fig:fig2}
\end{figure}

\noindent In Fig.~\ref{fig:fig2}, we show growth curve obtained in our
experiments at $P=1$~atm and $T=37^{\circ}$~C.
The growth curve shows a typical lag growth regime at small times
followed by an exponential growth phase and finally a saturation
regime. The value of the saturation OD $(<0.5)$ is smaller compared to
the saturation OD (typically 1.0) reached when oxygen is not a growth
limiting agent. In the oxygen limited environment both the division
time and saturation OD are affected.

\begin{figure}[!htb]
\begin{center}
\includegraphics[height=9cm,angle=270]{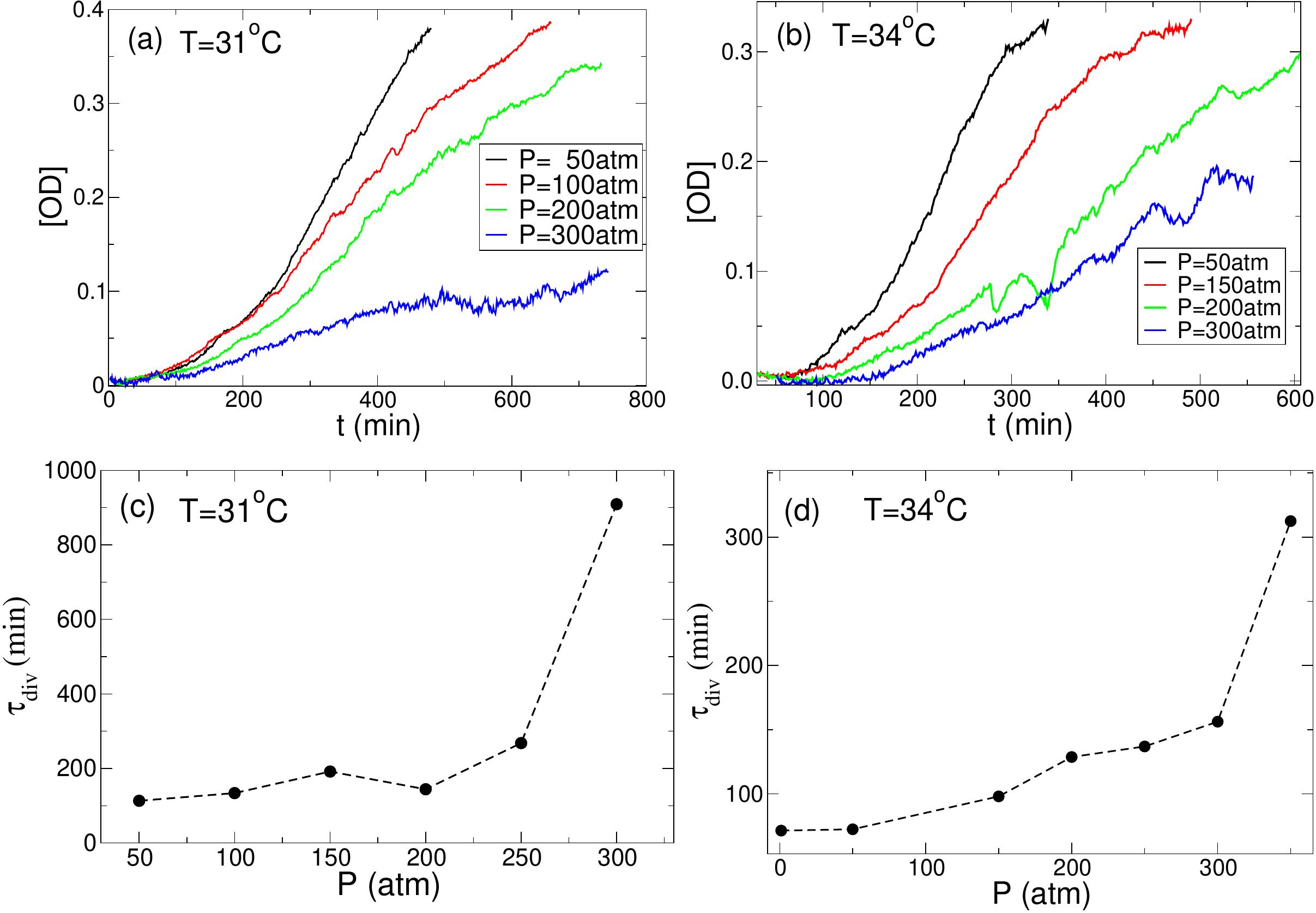}
\end{center}
\caption{(a) Growth curves at different pressures for
  $T=31^{\circ}$~C. (b) Growth curves at different pressures for
  $T=34^{\circ}$C. (c) Division (doubling) time $\tau_{\rm div}(P)$
  extracted from Fig.~\ref{fig:fig3}(a). (d) Division time $\tau_{\rm
    div}(P)$ extracted from Fig.~\ref{fig:fig3} (b). Pressure
  dependence of $\tau_{\rm div}(P)$ is marked by a sharp increase at
  high pressures where the cells still grow but the growth is
  extremely slow.}
\label{fig:fig3}
\end{figure}

\begin{figure}[!htb]
\begin{center}
\includegraphics[height=8cm]{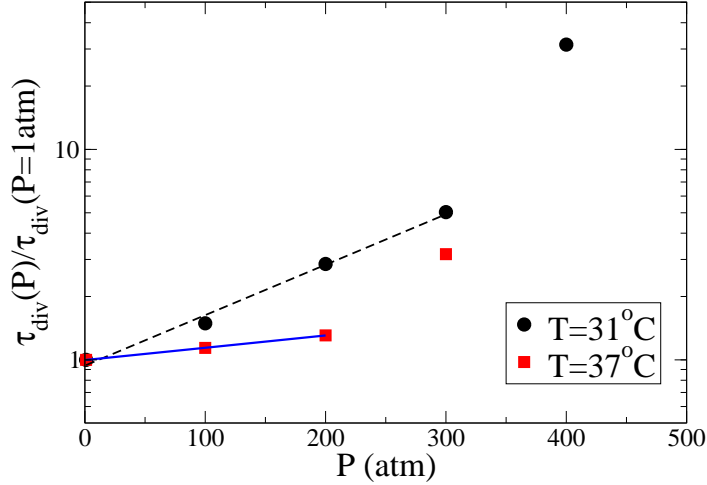}
\end{center}
\caption{Dependence of $\tau_{\rm div}(P)$ for two different
  temperatures $T=31^{\circ}$~C and $T=37^{\circ}$~C on a linear-log
  plot. The low pressure linear depedence on a linear-log plot
  suggests that $\tau_{\rm div}(P)$ follows an exponential
  behavior. The discontinous jump in $\tau_{\rm div}(P)$ at a given
  temperature is marked by its departure from the initial exponential
  behavior.}
\label{fig:fig4}
\end{figure}

\noindent In Figs.~\ref{fig:fig3} (a) and (b), we show the growth
curve of {\it E. coli} for various pressures at $T=31^{\circ}$C and
$T=34^{\circ}$~C respectively. We find that where the saturation is
reached within the time scale of our experiments, the time profile of
the growth curves show the typical characteristics of growth at
$P=1$~atm and $T=37^{\circ}$C. The number of bacterial cells at time
$t$ in the exponential regime can be written as:
\begin{equation}
N(t) = N(0).2^{t/\tau_{\rm div}}
\label{eq:eq2}
\end{equation}
where $N(0)$ is the number of bacterial cell at the beginning of the
exponential phase and $\tau_{\rm div}$ is the division
time. $t/\tau_{\rm div}$ corresponds to the number of generation in a
given time $t$. In Fig.~\ref{fig:fig3} (c) and (d), we show $\tau_{\rm
  div}$ extracted from Figs.~\ref{fig:fig3} (a) and (b) for various
pressures at $T=31^{\circ}$C and $T=34^{\circ}$C respectively. We find
that $\tau_{\rm div}(P)$ increases, and hence the rate of growth
decreases, upon increasing pressure. We further find that the $OD$
corresponding to the saturation regime decreases upon increasing
pressure. Earlier studies on the effect of pressure on the total
biomass production of different bacteria have found a similar decrease
in total mass as a function of pressure~\cite{Matsumura1974}.
The division time $\tau_{\rm div}(P)$ at a given temperature increases
with pressure but shows a discontinous jump at high pressures. We find
that the discontinuous jumps in $\tau_{\rm div}$ ocurrs between
$P=300-400$~atm for all the temperatures studied in our
experiments. To further characterize the low pressure regime of
$\tau_{\rm div}$, in Fig.~\ref{fig:fig4}, we show the division time
$\tau_{\rm div}(P)$ as a function of pressure for two different
temperature on a linear-log plot. We find that the low pressure regime
of increase of division time with pressure can be fit by an
exponential function where the exponent increases with decreasing
temperature. The discontinuous change in $\tau_{\rm div}(P)$ coincides
with departure from exponential behavior.

Pressure and temperature do not only affect the structural stability
of biomolecules but can also affect the thermodynamic force driving
different biochemical processes inside the cell. In general, the time
scale of a given chemical reaction is proportional to $e^{
  \frac{P\Delta V}{k_BT}}$, where $k_B$ is the Boltzmann constant and
$\Delta V$ is the volume change across the chemical reaction. It is
easy to see that any chemical reaction accompanied by a positive
volume change will exponentially slowdown with pressure. In this
context, the exponential dependence of $\tau_{\rm div}(P)$ with
pressure~(Fig.\ref{fig:fig4}) is not a surprise. Note that it is a
very simple consideration as most of the biochemical processes are not
invidividual but usually involve a cascade of chemical reacations
corresponding to any cellular module. Nonetheless, the exponential
dependence of the division time with pressure does suggest an overall
positive volume change. Furthermore, $\Delta V$ itself is a function
of pressure and temperature. At moderate pressures and temperature one
may assume it to be a constant.  It is hard to speculate the
mechanisms responsible for slow growth and further experiments must be
carried out to precisely figure out the decrease of growth rate at
high pressure.

The other remarkable feature of the pressure dependence of division
time is the abrupt increase of $\tau_{\rm div}(P)$ in the range of
pressures 200-400~atm for all the temperatures studied here. Where
does this discontinuity in the pressure dependence of growth come from
? Discontinuity in the growth as a function of pressure suggest that
something abrupt must happen in these range of pressure. The range of
pressures where we see a discontinous jump in the division time can
not be attributed to protein denaturation as the pressure is not high
enough to denature the proteins. While the proteins stability is
rather unaffected, the functionality of proteins may show a large
variability in this range of pressures~\cite{Merrin2011}. We
hypothesize that the discontinuous jump in the division time stem from
functional changes in biomolecules.

\section*{Pressure-temperature phase diagram of the division time of \textit{E. coli}}

\begin{figure}[!htb]
\begin{center}
\includegraphics[width=9cm]{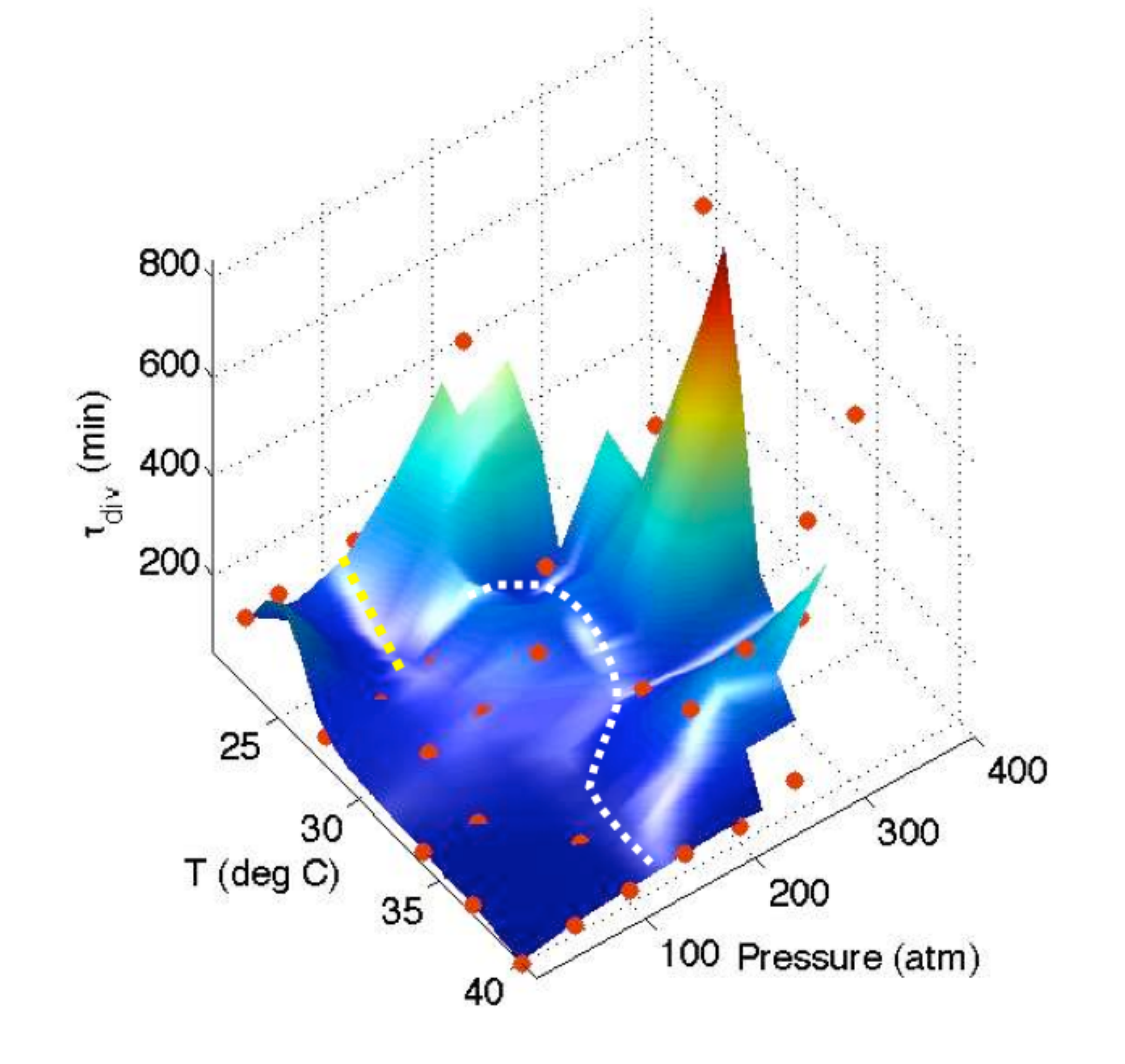} 
\end{center} 
\caption{Pressure temperature surface plot of division time $\tau_{\rm
    div} (P,T)$. Solid red circles are the experimental data
  points. White dotted line marks the loci of the points where
  $\tau_{\rm div}$ changes abruptly. Yellow dotted line with
  $\frac{dP}{dT}<0$ is region separating anomalous pressure dependence
  of the division time.}
\label{fig:fig5} 
\end{figure}

\noindent In Fig.~\ref{fig:fig5}, we show the surface plot of
pressure-temperature dependence of $\tau_{\rm div}(P,T)$. It shows
smooth change as a function of pressure and temperature but high
pressures as well as low temperatures growth are marked by sharp
change in $\tau_{\rm div}$. We further find that the slope of the
locus of the points in the (P,T) plane where $\tau_{\rm div}(P,T)$
shows sharp transition with respect to pressure resembles the
functional phase diagram of a typical protein (shown as dotted white
curve in Fig.~\ref{fig:fig5})~\cite{Merrin2011,Chilukuri2002}. A careful observation of the $\tau_{\rm
  div}(P,T)$ data reveals that at low $T$, there is a region where
$\tau_{\rm div}$ exhibits a non-monotonic behavior with pressure. In
this narrow region of pressure and temperature, $\tau_T(P)$ first
decreases and then increases further upon increasing pressure. The
yellow dotted line with $\frac{dP}{dT}>0$ marks the boundary between
this anomalous behavior and normal behavior of increasing division
time with increasing pressure. We hypothesize that this anomalous
behavior of division time as a function pressure results from
structural transition in the phospholipids present in the cell
membrane at low temperatures~\cite{Winter2005}.
\section*{Bacterial cell elongation, length distribution and heterogeneities:~~~}

\begin{figure}[!htb]
\begin{center}
  \includegraphics[width=9cm]{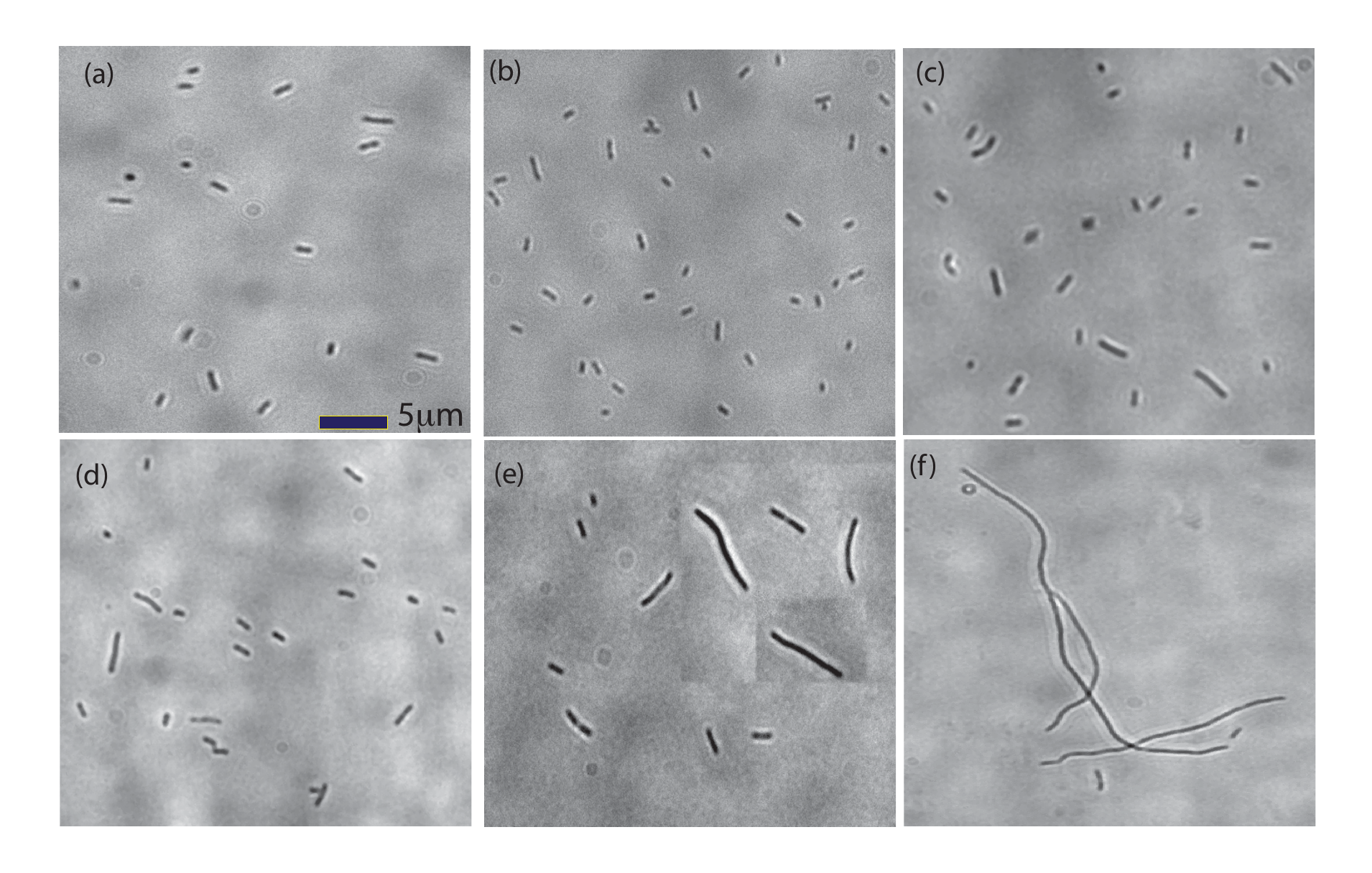}
\end{center}
\caption{Progressive elongation of bacterial cells at different
  pressures for $T=31^{\circ}$~C. (a) 1 atm (b) 50 atm (c) 100 atm (d)
  200 atm (e) 250 atm (f) 300 atm. The images were taken and
  analysized at the end of experiments for all the pressures.}
\label{fig:fig6}
\end{figure}

\begin{figure}[!htb]
\begin{center}
  \includegraphics[width=9cm]{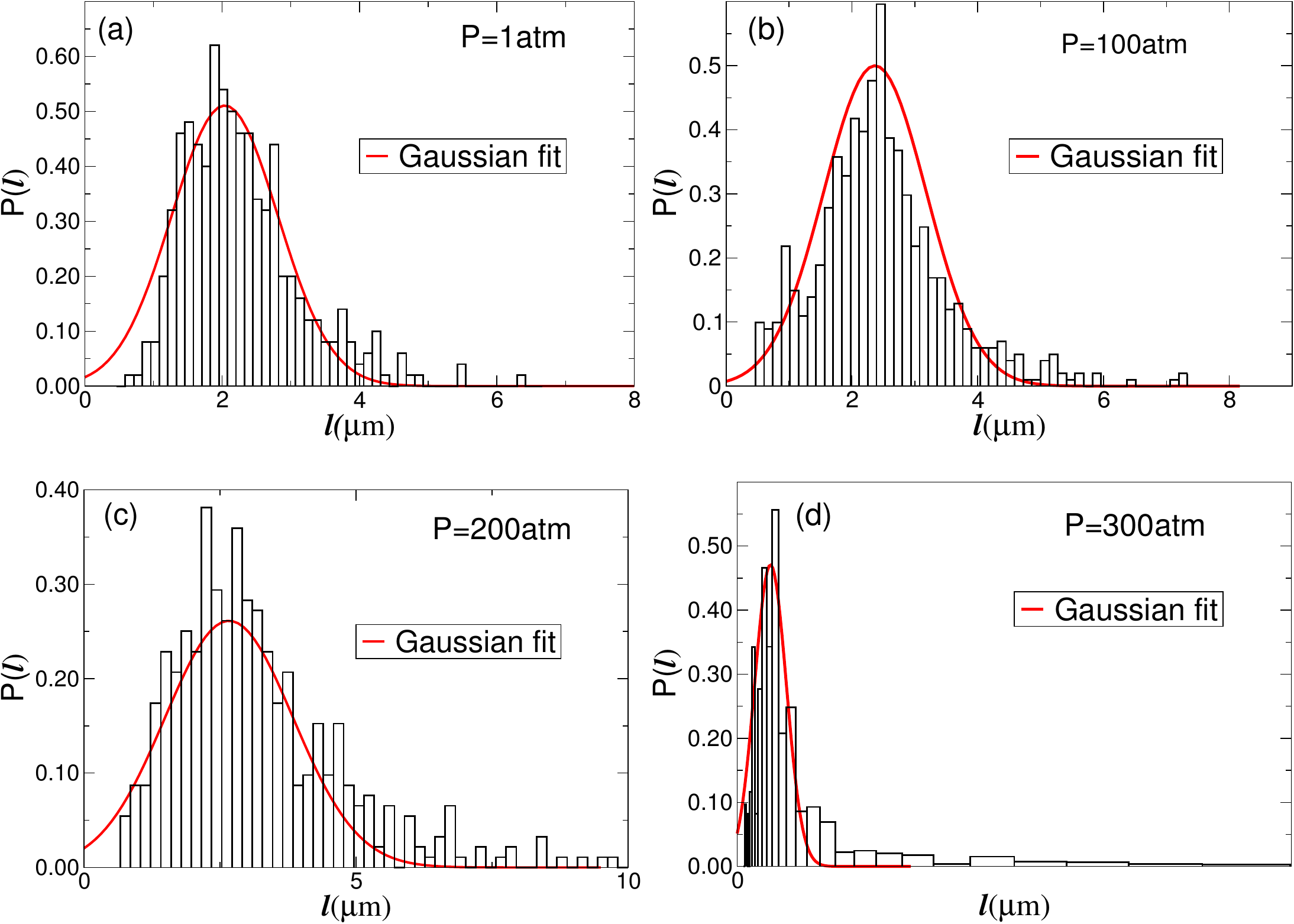}
\end{center}
\caption{Histogram of length of bacterial cells at different pressures
  for $T=31^{\circ}$~C. (a) $1$~atm (b) $100$~atm (c) $200$~atm (d)
  $300$~atm.}
\label{fig:fig7}
\end{figure}

\noindent Besides the slow growth of the population of bacterial cells
at high pressures, the other interesting features of the response to
high pressure is found in the morphological changes in bacterial
cells~\cite{Zobell1964,Ishii2004}. We find that average bacterial cell
length of E. Coli increases as a function of pressure (see
Fig.~\ref{fig:fig6}).
\noindent To further characterize the bacterial elongation, we looked
at the distribution of bacterial cell length at various pressures at a
given temperature. In Fig.~\ref{fig:fig7}, we show the distribution of
bacterial cell length $l$ at the end of our experiments for pressures
$P=1,100,200,\& 300$~atm respectively for $T=31^{\circ}$~C. The
distribution $P(l)$ of $l$ at $P=1$~atm follows a Gaussian
distribution. As the pressure is increased, $P(l)$ starts developing a
non-Gaussian tail suggesting a growing bacterial cell length
heterogeneities. A major fraction of the cells still retain the same
morphology as $P=1$~atm, but there is an increase in the population of
elongated cells upon increasing pressure.

\begin{figure}[!htb]
\begin{center}
  \includegraphics[width=8cm]{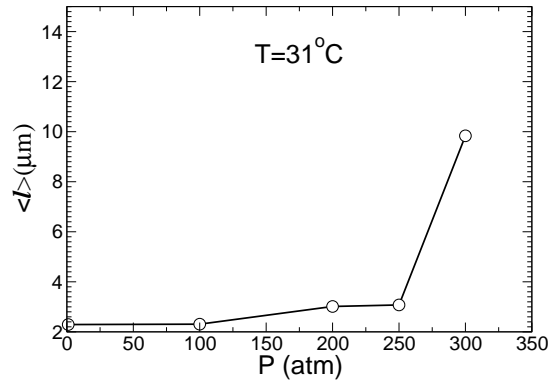}
\end{center}
\caption{Average bacterial cell length $\langle l \rangle $ as a
  function of pressure at $T=31^{\circ}$~C. Average length of
  bacterial cells shows a sharp transition between $P=250$ and
  $300$~atm.}
\label{fig:fig8}
\end{figure}

The average value of the cell length $\langle l \rangle$ increases
upon increasing pressure and shows a sharp increase at the same
pressure where the division time also shows a sharp increase (see
Fig.~\ref{fig:fig8}). While the bacterial cell elongation at high
pressure is known, the sharp transition at high pressure is
new. Furthermore, we find that the pressure and temperature where the
growth is marked by a sharp increase in division time correlates well
with sharp changes in bacterial cell length. The exponential increase
of division time with pressure as we saw in the earlier section can be
interpreted as exponential decrease of overall kinetics leading to
slow growth due to cell elongation at high pressure. While the
increased cell length upon increasing pressure would explain the
decreased rate of growth, it is not clear if the elongated cells would
grow slower than the cells with normal morphology. For example, if the
elongated cells just lack the ability of cell division but replicate
their genome normally then one would expect the growth rate per unit
cell size not to change unless other kinetic processes also get
affected by the increase of pressure.

\noindent The elongation of bacterial cells at high pressure has been
a subject of intense research and to our knowledge no consensus on the
molecular mechanism responsible for it is reached~\cite{Ishii2004}. To
account for the bacterial cell length heterogeneities and elongation
upon increased pressure, we propose a stochastic model in the next
section.

\section*{A stochastic irreversible switch model for the morphological changes at high pressures}

\noindent A quick overview of Figs.~\ref{fig:fig6} and ~\ref{fig:fig7}
suggests that while the average length of the bacterial cells
undergoes a sharp transition at high pressures, a major fraction of
bacterial still retain the morphology of a normal cell.
\begin{figure}[!htb]
\begin{center}
  \includegraphics[width=8cm,height=4cm]{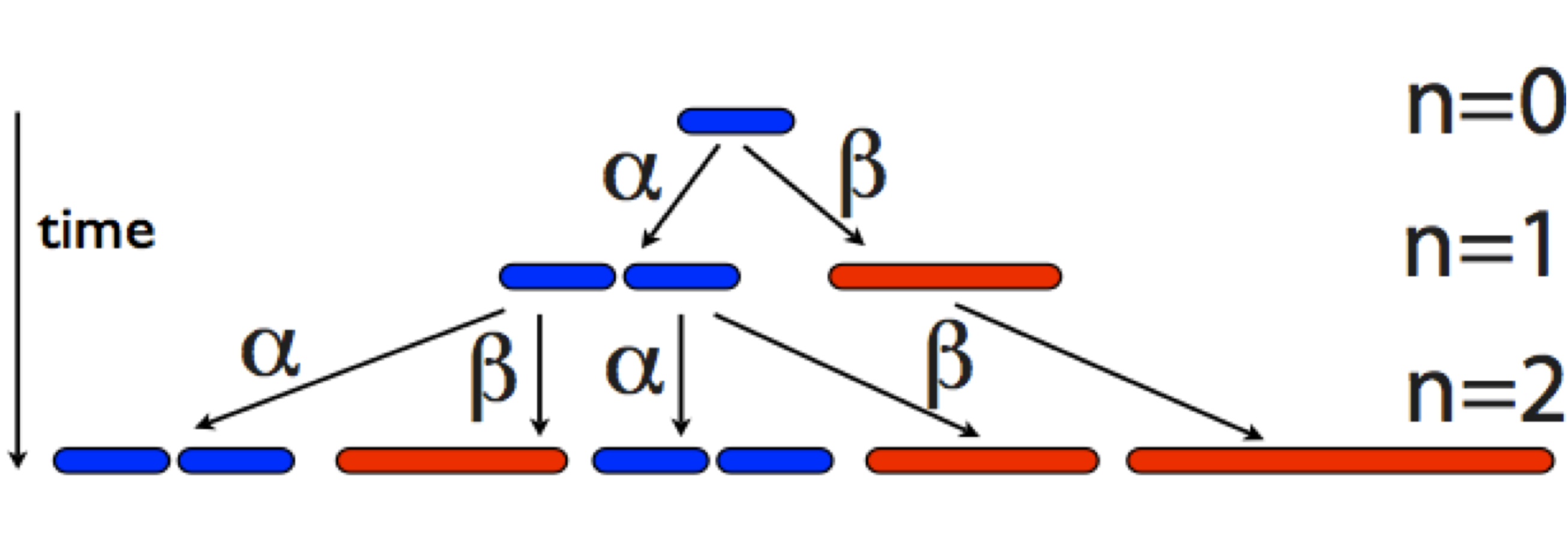}
\end{center}
\caption{Schematic of the stochastic irreversible switching of normal
  bacterial cell (blue) to filamenting bacterial cells (red). A normal
  cell can either divide into two cells with probability $\alpha$ or
  switch to a filamenting phenotype with a probability $\beta$. Once a
  bacterial cell's fate changes to filamentation, it just grows
  without dividing.}
\label{fig:fig9}
\end{figure}
The change of morphology can be thought of as an irreversible switch
of bacterial morphology during the course of growth of the bacterial
cell where the rate of switching will depend on the pressure. Let us
assume that we start with $N_0$ cells at time $t=0$. Let's further
assume that the cells either decide to divide into two identical cells
with probability $\alpha$ or grow irreversibly to a filamenting
bacteria with probability $\beta = (1-\alpha)$. Let's further assume
that the internal growth rate of both the normal cells and the
filamenting cells is the same and hence whenever bacterial do not
divide from one generation $n$ to a generation $n+1$, the cell length
just doubles (see Fig.~\ref{fig:fig9}). Hence at the end of the $n$
generations of division, the system will have different distribution
of bacterial cell length $l \in \{l_0, 2l_0, 4l_0, 8l_0,
....,2^nl_0\}$. We can assume the initial distribution of bacterial
length, as suggested by our experimental data, to be a Gaussian given
by:
\begin{equation}
P(l, t=0)= \frac{1}{\sqrt{2\pi\sigma_l^2}}e^{-(l-l_0)^2/\sigma_l^2}
\end{equation}
It can be shown easily that the above scheme of irreversible
stochastic switching leads to number of various lengths $l$ of
bacterial cells at the end of $n$ generations given by:
\begin{eqnarray*}
N(l=l_0) &=& (2\alpha)^n \\
N(l=2l_0) &=& (2\alpha)^{(n-1)}\beta\\
N(l=4l_0) &=& (2\alpha)^{(n-2)}\beta^2\\
N(l=8l_0) &=& (2\alpha)^{(n-3)}\beta^3\\
\end{eqnarray*}
Hence in the extreme cases (i) no switching or $\alpha=1$ will lead to
no changes in the bacterial length (ii) $\alpha=0$, all the cells will
elongate to the maximum limited by growth and number of divisions. In
general, the number of bacterial cells of length $l=2^al_0$ is given
by:
\begin{equation}
N(l=2^al_0) = (2\alpha)^{(n-a)}\beta^{a}
\end{equation}
The total number of bacterial cells at the end of $n$ generation can be given by:
\begin{equation}
N = \sum_{a=0}^{n} (2\alpha)^{(n-a)}\beta^{a}=(2\alpha)^n \sum_{a=0}^{n}(\frac{\beta}{2\alpha})^a = (2\alpha)^n\frac{1-(\beta/2\alpha)^{n+1}}{1-\beta/2\alpha}
\end{equation}
Now the probability $P(l=2^al_0)$ of a bacterial cell with length $2^al_0$
is given by:
\begin{equation}
P(l=2^al_0) = \frac{N(l=2^al_0)}{N} = \frac{(2\alpha)^{(n-a)}\beta^{a}}{(2\alpha)^n\frac{1-(\beta/2\alpha)^{n+1}}{1-\beta/2\alpha}}
\end{equation}
which in terms of the switching probability $\beta$ can be written as:
\begin{equation}
P(l=2^al_0) = (\frac{\beta}{2(1-\beta)})^a\frac{1-\frac{3}{2}\beta}{(1-\beta)[1-(\frac{\beta}{2(1-\beta)})^{n+1}]}
\end{equation}
Hence the expectation value of length $\langle l_n \rangle$ at the end
of $n$ generation is given by:
\begin{eqnarray*}
\langle l_n \rangle = \sum_{a=0}^{n}2^al_0.P(l=2^al_0) &=& l_0\frac{1-\frac{3}{2}\beta}{(1-\beta)[1-(\frac{\beta}{2(1-\beta)})^{n+1}]}\sum_{a=0}^{n}(\frac{\beta}{1-\beta})^a \\
                                                     &=& \frac{(1-\frac{3}{2}\beta)}{(1-2\beta)}\frac{[1-(\frac{\beta}{1-\beta})^{n+1}]}{[1-(\frac{\beta}{2(1-\beta)})^{n+1}]}
\end{eqnarray*}
Since the distribution of $l$ at $t=0$ is a Gaussian, the distribution
of length at the end of $n$ generations can be written as:
\begin{eqnarray}\nonumber
P_n(l) &=& \sum_{a=0}^{n}(\frac{\beta}{2(1-\beta)})^a\frac{1-\frac{3}{2}\beta}{(1-\beta)[1-(\frac{\beta}{2(1-\beta)})^{n+1}]}  \\ 
&&  \frac{1}{\sqrt{2\pi\sigma_l^2}2^{2a}}e^{-(l-2^a.l_0)^2/(2^{2a}\sigma_0^2)}
\label{eq:pnofl}
\end{eqnarray}

\begin{figure}[!htb]
  \begin{center}
    \includegraphics[width=9cm]{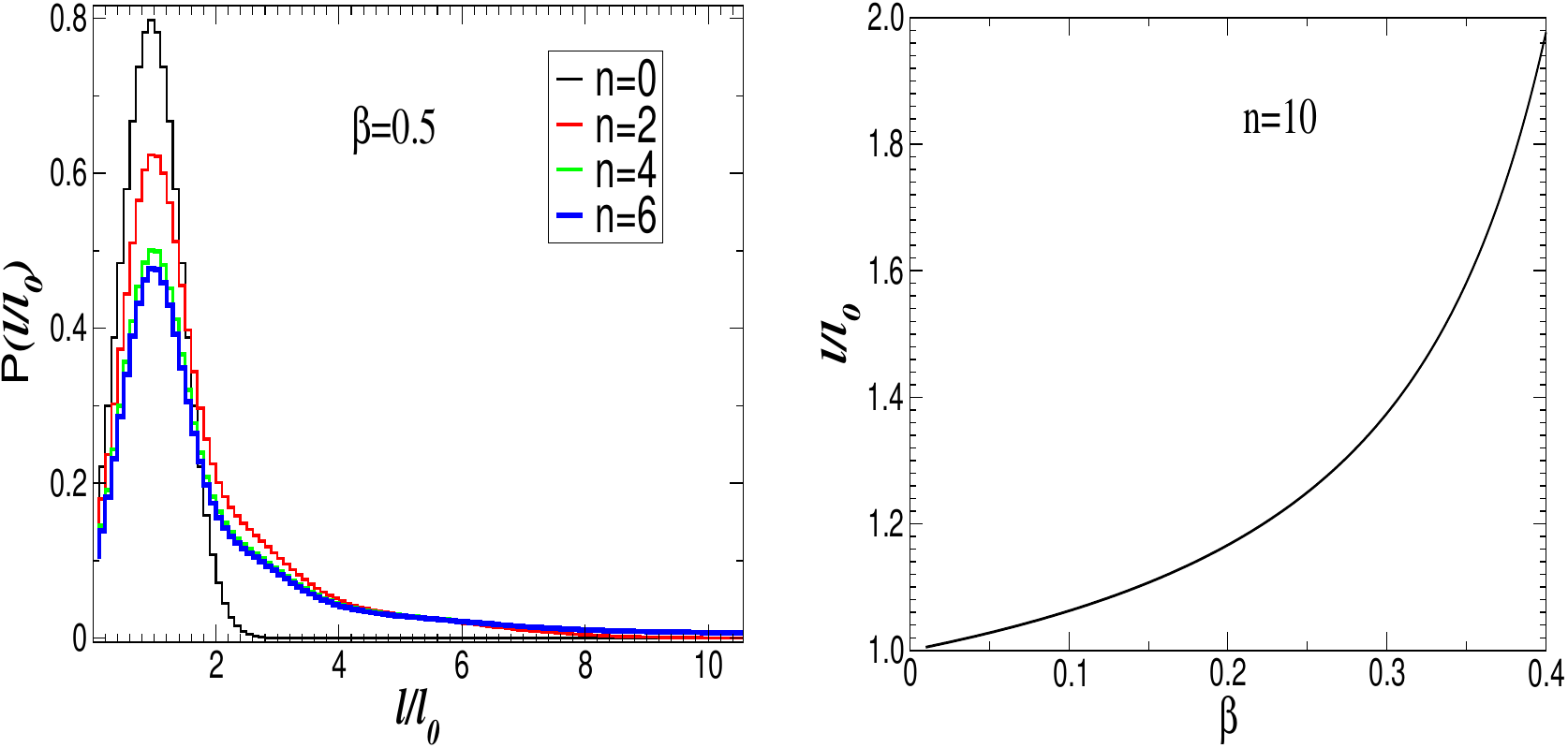}
  \end{center}
\caption{(a) Evolution of the distribution of the length of bacterial
  cells for $\beta=0.5$. (b) Model prediction of the average length
  as a function of switching probability $\beta$.}

\label{fig:lengthvsbeta}
\end{figure}
\begin{figure}[!htb]
  \begin{center}
    \includegraphics[width=9cm]{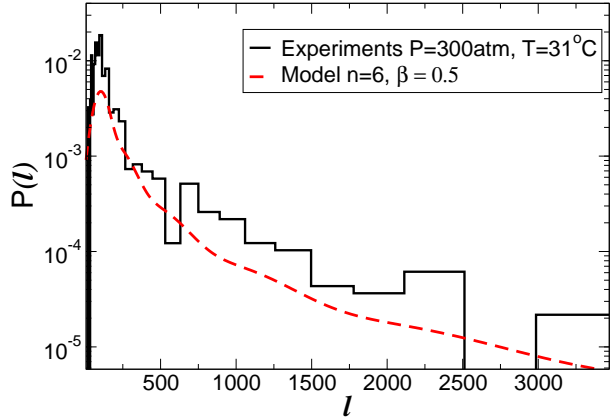}
  \end{center}
  \caption{Comparison of model prediction of $P_n(l)$ with
    experimental data at $P=300$~atm and $T=31^{\circ}$~C with model
    parameters $\beta=0.5$ and $n=6$. Experimental data is shown in
    solid black line while the model prediction is shown in dashed red
    line.}
  \label{fig:distexpmodel}
\end{figure}

\noindent In Fig.~\ref{fig:lengthvsbeta}(a), we show the evolution of
the distribution of $l$ for a fixed value of $\beta=0.5$. As the time
progresses the distribution develops a long tail. Note that for $\beta
> \beta_c=2/3$, the system would undergo an irreversible fate where
after few generations the population will be dominated by elongated
cells and cells with normal length will vanish from the population in
the limit of long time. We further show the expectation value of
length $\langle l \rangle$ in Fig.~\ref{fig:lengthvsbeta} (b) as a
function of $\beta$ for a fixed number of generations. $\langle l
\rangle$ increases slowly for small $\beta$ values and growth sharply
with increasing $\beta$. To further test our model, we compare the
data of $P(l)$ at $P=300$~atm with our model in
Fig.~\ref{fig:distexpmodel}. We find that the model can reasonably
reproduce the length distribution.
To further characterize the heterogeneities in the
population of bacterial cell length, we calculate a non-Gaussian
measure $\phi$~\cite{Rahman1964} of the distribution $P_n(l)$defined by:
\begin{equation}
\phi = \frac{\langle \Delta l^4 \rangle}{3(\langle \Delta l^2 \rangle)^2}-1
\end{equation}
where $\langle \Delta l^2 \rangle$ and $\langle \Delta l^4 \rangle$
are the second and fourth central moments of the distribution $P_n(l)$
respectively. $\phi=0$ corresponds to a Gaussian distribution, while a
deviation of $\phi$ from zero corresponds to the degree of deviation
from a Gaussian distribution.

\begin{figure}[!htb]
  \begin{center}
    \includegraphics[width=8cm]{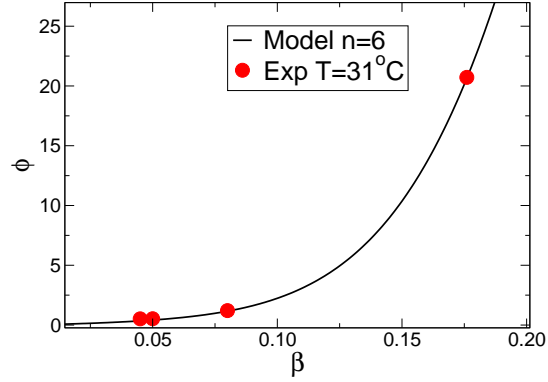}
  \end{center}
\caption{Dependence of non-Gaussianity of the distribution $\phi$ on
  the switching probability $\beta$. Model data is shown in solid
  black curve. We have also plotted the values of $\phi$ extracted
  from experimental distribution of cell length at $T=31^{\circ}$~C
  for pressures $P=50,100,150,$ and $200$~atm as solid red circles.}
  \label{fig:phivsbeta}
\end{figure}

\noindent In Fig.~\ref{fig:phivsbeta}, we show the dependence of
$\phi$ on switching probability $\beta$ for $n=6$. We find that $\phi$
grows slowly first but increases sharply with $\beta$. In
Fig.~\ref{fig:phivsbeta}, we also show the non-Gaussianity parameter
$\phi$ measured from the experimental distribution of cell lengths at
pressures $P=50,100,150,$ and $200$~atm and temperature
$T=31^{\circ}$~C as solid red circles. Note that model assumes a
transition but is not able to say much about the physical origin of
such phenotypic transition. What are the biophysical mechanisms
responsible for the cell elongation ? Where does the stochasticity
come from ? The clue to the latter comes from the measured transition
in the cell length observed here and the polymerization of one of the
cytoskeletal proteins responsible for cell division, FtsZ. Recent
experiments on FtsZ {\it in vivo} and {\it in vitro} suggests that
FtsZ protein depolymerizes at high pressures leading to delocalization
of FtsZ in the cell~\cite{Ishii2004}. Furthermore, it was shown that
FtsZ is not able to form the Z-ring which is considered responsible
for the mechanical forces required for the cell division. Could FtsZ
be responsible for the sharp transition in the growth and the cell
division observed in our experiments ? Or a set of other processes,
including the formation of Z-ring by FtsZ, lead to the observed
transition ? Is the cell elongation phenomenon due to only the
depolymerization of FtsZ at high pressures or more than one cellular
processes are responsible for it ?  The answers to all these questions
can only come from further experiments that we are performing.

\section*{Summary and Discussion}

\noindent We have investigated the growth of \textit{E. coli} in real
time as a function of pressure and temperature. We find that {\it
  E. coli} can grow and divide in a wide range of pressures
($1-400$~atm) and temperatures $(20-40^{\circ}C)$. The division time
of bacteria increases upon increasing pressure at a given
temperature. Furthermore, division time at a constant temperature
exhibits an exponential dependence on pressure for moderate values of
pressure. Moreover, we find that for all the temperatures studied,
division time shows an abrupt increase at pressures between
$250-400$~atm. While at high $T$ this sharp increase in division time
with pressure is very large where $\tau_{div}$ can be larger than
$500$~min, at low $T$, $\tau_{div}$ increases by few
folds. Furthermore, we find that the division time shows an anomalous
decrease and then increase with pressure at low temperature. We
hypothesize that this anomalous behavior of division time is a
manifestation of the structural changes in phospholipids in the
membrane. Further experiments on a veriety of cell types where the
lipid composition is known would be required to answer this question.

\noindent We next looked at the bacterial cell morphology after
application of pressure till the time of saturation in the cases where
we could reach the saturation or few generation times where the
saturation was hard to reach over the time scale of our
experiments. We find that average bacterial length increases upon
pressure. While the bacterial elongation at high
pressures~\cite{Zobell1964,Ishii2004} is known, we find that
E.Coli. shows a behavior of morphology very similar to growth rate or
division time, whereas the average cell length also displays a sharp
increase at pressures between $250$ and $400$~atm. Moreover, the
heterogeneities in the cell length of bacteria increases upon
increasing pressure. To explain the heterogeneities in the cell length
with pressure, we propose a simple stochastic irreversible switch
model of bacterial phenotypes (normal and filamenting). We find that
the model fits well the experimental data of distribution of bacterial
cell length at different pressure. Moreover, the model allows us to
extract the switching probability of E. Coli. bacteria to filamenting
phenotype, which increases upon increasing pressure. While the model
captures the cell elongation phenomenon and explains the cell length
distribution, it leaves us with many questions such as -- what
biophysical processes give rise to the stochasticity in the phenotypic
transitions as a function of pressure ? A clue to this comes from the
measured transition in the cell elongation observed here and
depolymerization of FtsZ protein responsible for cell division. Note
that since the bacterial strain used in our experiments (DH$5\alpha$)
lacks the homologous recombination system, the cell elongation can no
be interpreted as the conventional SOS response of the system. Hence
FtsZ depolymerization and delocalization leading to non-formation of a
Z-ring is a potential biophysical process that may lead to phenotypic
transitions proposed here. Further experiments are required where the
polymerization of cytoskeletal proteins such as MreB and FtsZ can be
visualized along with cell division at various pressures and
temperatures.

\noindent Since growth is coupled to various other processes the
bottlenecks could be either the structural integrity (such as protein
denaturation or membrane structural changes) or the time integrity of
various processes. There is a large body of literature on the behavior
of different biomolecules at varied physical conditions. These studies
indicate that at high pressures and temperatures the essential
components that make up a cell may become unstable. Proteins can
unfold and membranes can undergo structural transitions at high
pressures leading to death of a
cell~\cite{Jaenicke1998,Razvi2006}. The other issue which has been
rather overlooked in past is the variation in the time scales of
various processes. Since pressure and temperature not only change the
stability but they also modify the thermodynamic driving force of a
chemical reaction and hence lead to changes in time scales of various
processes. How the time integrity of various processes is maintained
by a cell is an interesting question. A better understanding can only
come from a systematic study of the mutations in the protein/enzymes
or regulatory circuits involved in various processes.


\begin{addendum}
 \item We thank A. Buguin, Y. T. Maeda and J. Merrin, and H. E. Stanley for helpful
   discussions. AL acknowledges Florence Gould fellowship from the
   Institute for Advanced Study, Princeton, NJ USA and NSF grant
   No. PHY-0848815 for support.

 \item[Competing Interests] The authors declare that they have no
competing financial interests.
 \item[Correspondence] Correspondence and requests for materials
should be addressed to Pradeep Kumar~(email: pradeep.kumar@rockefeller.edu).
\end{addendum}

\end{document}